\documentclass[a4paper,12pt]{article}
\usepackage[a4paper,inner=30mm,outer=30mm,top=35mm,bottom=25mm]{geometry}
\usepackage{pdflscape}
\usepackage{afterpage}
\usepackage{multirow}
\usepackage{amsmath}
\usepackage{graphicx}

\newcommand{\ba}{\begin{eqnarray}}
\newcommand{\ea}{\end{eqnarray}}
\newcommand{\arry}[2]{\begin{array}{#1} #2 \end{array}}
\begin{document}

\begin{titlepage}
\thispagestyle{plain}
\pagestyle{plain}      
\rightline{LTH-1107} 
\vspace{1.5cm}

\begin{center}
 {\Large \bf 
Niemeier Lattices in the \\Free Fermionic Heterotic--String 
formulation 
}
\end{center}

\begin{center}
{\large
Panos Athanasopoulos$^*$ and
Alon E. Faraggi$^*$
}
\\

\bigskip

{$^*$} {Department of Mathematical Sciences,\\
		University of Liverpool,     \\
                Liverpool L69 7ZL, United Kingdom}
\end{center}

\smallskip

\begin{abstract}

The spinor--vector duality was discovered in 
free fermionic constructions of the heterotic-string in four 
dimensions. It played a key role in the 
construction of heterotic--string models with an anomaly free 
extra $Z^\prime$ symmetry that may remain unbroken down to low 
energy scales. A generic signature of the low scale string
derived $Z^\prime$ model is via di--photon excess that may 
be within reach of the LHC. A fascinating possibility is 
that the spinor--vector duality symmetry is rooted in the 
structure of the heterotic--string compactifications to two
dimensions. The two dimensional heterotic--string theories 
are in turn related to the so--called moonshine symmetries 
that underlie the two dimensional compactifications. In 
this paper we embark on exploration of this connection by
the free fermionic formulation to classify the symmetries of the 
two dimensional heterotic--string theories. We use two complementary
approaches in our classification. The first utilises a construction 
which is akin to the one used in the spinor--vector duality. Underlying
this method is the triality property of $SO(8)$ representations. 
In the second approach we use the free fermionic tools to 
classify the twenty four dimensional Niemeier lattices. 

\end{abstract}

\end{titlepage}
\setcounter{page}{2}
\def\beq{\begin{equation}}
\def\eeq{\end{equation}}
\def\beqn{\begin{eqnarray}}
\def\eeqn{\end{eqnarray}}

\def\no{\noindent }
\def\nolabel{\nonumber }
\def\ie{{\it i.e.}}
\def\eg{{\it e.g.}}
\def\half{{\textstyle{1\over 2}}}
\def\third{{\textstyle {1\over3}}}
\def\quarter{{\textstyle {1\over4}}}
\def\sixth{{\textstyle {1\over6}}}
\def\m{{\tt -}}
\def\p{{\tt +}}

\def\Tr{{\rm Tr}\, }
\def\tr{{\rm tr}\, }

\def\slash#1{#1\hskip-6pt/\hskip6pt}
\def\slk{\slash{k}}
\def\GeV{\,{\rm GeV}}
\def\TeV{\,{\rm TeV}}
\def\y{\,{\rm y}}
\def\SM{Standard--Model }
\def\SUSY{supersymmetry }
\def\SSSM{supersymmetric standard model}
\def\vev#1{\left\langle #1\right\rangle}
\def\l{\langle}
\def\r{\rangle}
\def\o#1{\frac{1}{#1}}

\def\Htw{{\tilde H}}
\def\chibar{{\overline{\chi}}}
\def\qbar{{\overline{q}}}
\def\ibar{{\overline{\imath}}}
\def\jbar{{\overline{\jmath}}}
\def\Hbar{{\overline{H}}}
\def\Qbar{{\overline{Q}}}
\def\abar{{\overline{a}}}
\def\alphabar{{\overline{\alpha}}}
\def\betabar{{\overline{\beta}}}
\def\tautwo{{ \tau_2 }}
\def\thetatwo{{ \vartheta_2 }}
\def\thetathree{{ \vartheta_3 }}
\def\thetafour{{ \vartheta_4 }}
\def\ttwo{{\vartheta_2}}
\def\tthree{{\vartheta_3}}
\def\tfour{{\vartheta_4}}
\def\ti{{\vartheta_i}}
\def\tj{{\vartheta_j}}
\def\tk{{\vartheta_k}}
\def\calF{{\cal F}}
\def\smallmatrix#1#2#3#4{{ {{#1}~{#2}\choose{#3}~{#4}} }}
\def\ab{{\alpha\beta}}
\def\Minv{{ (M^{-1}_\ab)_{ij} }}
\def\bone{{\bf 1}}
\def\ii{{(i)}}
\def\V{{\bf V}}
\def\N{{\bf N}}

\def\b{{\bf b}}
\def\S{{\bf S}}
\def\X{{\bf X}}
\def\I{{\bf I}}
\def\mb{{\mathbf b}}
\def\mS{{\mathbf S}}
\def\mX{{\mathbf X}}
\def\mI{{\mathbf I}}
\def\balpha{{\mathbf \alpha}}
\def\bbeta{{\mathbf \beta}}
\def\bgamma{{\mathbf \gamma}}
\def\bxi{{\mathbf \xi}}

\def\t#1#2{{ \Theta\left\lbrack \matrix{ {#1}\cr {#2}\cr }\right\rbrack }}
\def\C#1#2{{ C\left\lbrack \matrix{ {#1}\cr {#2}\cr }\right\rbrack }}
\def\tp#1#2{{ \Theta'\left\lbrack \matrix{ {#1}\cr {#2}\cr }\right\rbrack }}
\def\tpp#1#2{{ \Theta''\left\lbrack \matrix{ {#1}\cr {#2}\cr }\right\rbrack }}
\def\l{\langle}
\def\r{\rangle}
\newcommand{\cc}[2]{c{#1\atopwithdelims[]#2}}
\newcommand{\bth}[2]{\bar\theta{#1\atopwithdelims[]#2}}
\newcommand{\nn}{\nonumber}


\def\inbar{\,\vrule height1.5ex width.4pt depth0pt}

\def\IC{\relax\hbox{$\inbar\kern-.3em{\rm C}$}}
\def\IQ{\relax\hbox{$\inbar\kern-.3em{\rm Q}$}}
\def\IR{\relax{\rm I\kern-.18em R}}
 \font\cmss=cmss10 \font\cmsss=cmss10 at 7pt
\def\IZ{\relax\ifmmode\mathchoice
 {\hbox{\cmss Z\kern-.4em Z}}{\hbox{\cmss Z\kern-.4em Z}}
 {\lower.9pt\hbox{\cmsss Z\kern-.4em Z}}
 {\lower1.2pt\hbox{\cmsss Z\kern-.4em Z}}\else{\cmss Z\kern-.4em Z}\fi}

\def\AEF{A.E. Faraggi}
\def\JHEP#1#2#3{{\it JHEP}\/ {\bf #1} (#2) #3}
\def\NPB#1#2#3{{\it Nucl.\ Phys.}\/ {\bf B#1} (#2) #3}
\def\PLB#1#2#3{{\it Phys.\ Lett.}\/ {\bf B#1} (#2) #3}
\def\PRD#1#2#3{{\it Phys.\ Rev.}\/ {\bf D#1} (#2) #3}
\def\PRL#1#2#3{{\it Phys.\ Rev.\ Lett.}\/ {\bf #1} (#2) #3}
\def\PRT#1#2#3{{\it Phys.\ Rep.}\/ {\bf#1} (#2) #3}
\def\MODA#1#2#3{{\it Mod.\ Phys.\ Lett.}\/ {\bf A#1} (#2) #3}
\def\IJMP#1#2#3{{\it Int.\ J.\ Mod.\ Phys.}\/ {\bf A#1} (#2) #3}
\def\nuvc#1#2#3{{\it Nuovo Cimento}\/ {\bf #1A} (#2) #3}
\def\RPP#1#2#3{{\it Rept.\ Prog.\ Phys.}\/ {\bf #1} (#2) #3}
\def\EJP#1#2#3{{\it Eur.\ Phys.\ Jour.}\/ {\bf C#1} (#2) #3}
\def\etal{{\it et al\/}}

\hyphenation{su-per-sym-met-ric non-su-per-sym-met-ric}
\hyphenation{space-time-super-sym-met-ric}
\hyphenation{mod-u-lar mod-u-lar--in-var-i-ant}


\setcounter{footnote}{0}
\section{Introduction}

\begin{figure}[!b]
\begin{center}
\includegraphics[width=10cm]{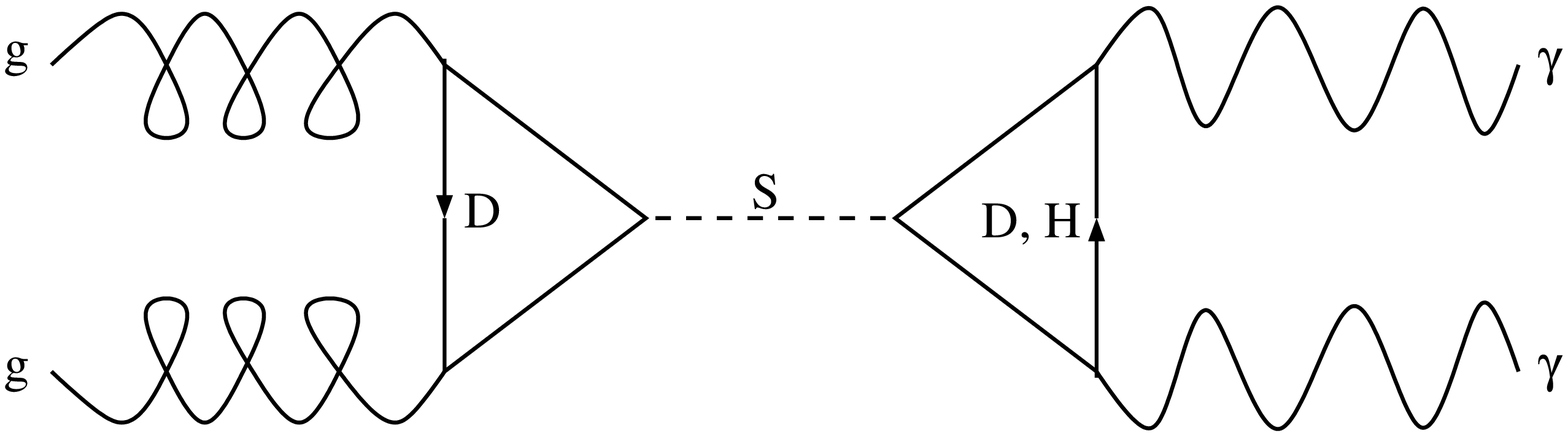}
\caption{Production and di--photon decay of the Standard Model 
singlet scalar state.}
\label{didiagram}
\end{center}
\end{figure}

The ATLAS and CMS collaborations reported in December 2015 evidence
for excess in the di--photon chanel 
\cite{atlasdec,cmsdec}. Absence of evidence for any other 
deviation from the Standard Model expected signals suggested 
that the excess could be interpreted as production and decay 
of a Standard Model singlet state by heavy vector--like states 
\cite{strumia}, in a process depicted in figure \ref{didiagram}.
In ref. \cite{frdiphoton} it was shown that the spectrum required to 
generate the excess naturally arise in the string derived 
model of ref. \cite{frzprime}, which allows for a light 
$Z^\prime$ vector boson. Anomaly cancellation mandates that 
the mass scale of the Standard 
Model singlet state, which is produced in resonance in figure 
\ref{didiagram}, as well as the mass scale of the heavy
vector--like states that are used in the production and decay 
of the singlet states, is the $Z^\prime$ symmetry
breaking scale. Thus, assuming that the $Z^\prime$ remains 
unbroken down to the multi--TeV scale naturally gives rise 
to the characteristics required to generate di--photon excess. 
In ref. \cite{adfm} it was shown that existence of the 
light $Z^\prime$ at the multi--TeV scale is compatible 
with gauge coupling unification at the GUT scale, as well
as other phenomenological constraints. 

In August 2016 the ATLAS and CMS collaborations reported that
accumulation of further data did not substantiate the 
observation of the di--photon excess \cite{atlasaug,cmsaug}, 
suggesting that initial observation was a statistical fluctuation. 
However, this does not repudiate the di--photon excess as a signal 
of the string derived $Z^\prime$ model, albeit not as the purported 
750GeV resonance. Thus, searching for di--photon excesses in the 
energy range accessible at the LHC continues to be of immense interest.

Extra $Z^\prime$ vector bosons as possible signatures of heterotic--string 
vacua have been discussed in the literature since the mid--eighties
\cite{zpreview}. 
The difficulty in constructing heterotic--string models that allow
for an extra $U(1)$ symmetry to remain unbroken down to low scales
stems from the fact that the aforementioned symmetries tend to be
anomalous in the heterotic--string derived models.
The reason being that the string models utilise the 
symmetry breaking pattern $E_6\rightarrow SO(10)\times U(1)_\zeta$, 
with anomalous $U(1)_\zeta$. Suppression of left--handed neutrino
masses implies that the extra $U(1)$ symmetry, which is 
embedded in $SO(10)$, has to be broken near the GUT scale. 
This conundrum motivated the search of extra $U(1)$ symmetries 
that do not admit the $E_6$ embedding of their charges \cite{none6zp}. 
However, these choices result in contradiction between gauge 
coupling unification and the gauge coupling parameters at the
electroweak scale, which works well if the extra $U(1)$ charges
admit the $E_6$ embedding \cite{zpgcu, adfm}. 

It is therefore notable that in reference \cite{frzprime}
an heterotic--string derived model with an anomaly free 
$U(1)_\zeta$ was constructed. What is perhaps more remarkable 
is that the construction of the model utilises a basic duality
symmetry that operates in the space of $Z_2\times Z_2$ 
heterotic--string vacua which was dubbed spinor--vector
duality. The duality operates under the exchange of the 
total number of spinorial $16\oplus\overline{16}$ and
vectorial $10$ representations of $SO(10)$. For 
every vacuum with a number of $16\oplus\overline{16}$
representations, and a number of $10$ representations, 
there exist a dual vacuum in which the two numbers are
interchanged. One can further show that the duality 
arises from the breaking of $(2,2)$ world--sheet supersymmetry
to $(2,0)$ and that the duality map is induced
by a spectral flow operator that operates in the 
bosonic sector of the heterotic--string vacuum. 
In the vacua with $(2,2)$ world--sheet supersymmetry 
the $SO(10)\times U(1)_\zeta$ symmetry is enhanced to $E_6$. 
The chiral $27$ and $\overline{27}$ representations of 
$E_6$ decompose under $SO(10)\times U(1)_\zeta$ as 
\beqn
27& = & 16_{+\frac{1}{2}}~~+~~10_{-1}~~+1_{+2},\nonumber\\
\overline{27}& = & \overline{16}_{-\frac{1}{2}}~~+~~10_{+1}~~+1_{-2}.\nonumber
\eeqn
Thus, the $(2,2)$ vacua are self--dual under the exchange 
of the total number of spinorial $16\oplus\overline{16}$ 
and vectorial $10$ representations. 
The spectral flow operator acts as the $U(1)$ generator of
the $N=2$ world--sheet supersymmetry and interchanges between
the $SO(10)$ components in the decomposition of 
$E_6$ under $SO(10)\times U(1)_\zeta$. 
The breaking of the 
$E_6$ to $SO(10)\times U(1)_\zeta$, or the breaking of the 
world--sheet supersymmetry from $(2,2)$ to $(2,0)$ is induced
by Wilson lines. One choice of Wilson line breaking results 
in a vacuum with 
$\#_1$ of $16\oplus\overline{16}$
representations, and a $\#_2$ of $10$ representations, 
whereas a second choice interchanges the two numbers. 
Furthermore, the duality map between the dual cases
induced by the spectral flow operator of the parent $(2,2)$
vacuum. 

A new twist is that the spinor--vector duality was
used to construct the heterotic--string model with 
anomaly free $U(1)_\zeta$ that allows for an extra $E_6$
$Z^\prime$ to remain unbroken down to low scales. Using the 
methods developed in refs. \cite{gkr,fknr,fkr,acfkr,frs}
for the classification of free fermionic models, 
a self--dual model under the spinor--vector duality is
fished from the landscape of vacua. The unbroken gauge symmetry
at the string level is $SO(10)\times U(1)_\zeta$, but the 
spectrum is self--dual under the exchange of the 
total number of spinorial 
$16\oplus\overline{16}$ 
and vectorial $10$ representations. Thus, the spectrum still forms
complete $E_6$ multiplets and consequently $U(1)_\zeta$ is anomaly free. 
This is possible in the $Z_2\times Z_2$ orbifold
if the different spinorial and vectorial components are 
obtained from different fixed points. Conversely, obtaining 
both the spinorial and vectorial representations at the same
fixed point necessarily implies that the gauge symmetry is enhanced to
$E_6$.  in the model of ref. \cite{frzprime} the $SO(10)$ symmetry is broken
at the string level to $SO(6)\times SO(4)$. However, the chiral spectrum 
of the model still appears in complete $E_6$ representations,
hence maintaining $U(1)_\zeta$ as an anomaly free symmetry. 

The spinor--vector duality is a fundamental
symmetry in the space of $(2,0)$ heterotic--string vacua.
It played a central role in the construction of the 
$Z^\prime$ model in ref. \cite{frzprime}.
If the additional
$U(1)$ symmetry remains unbroken down to low scales
it may be detected via di--photon production as in fig. \ref{didiagram}. 

Another fascinating direction of investigation is the 
possibility that the spinor--vector duality is a mere reflection 
of a much larger symmetry structure that underlies this
class of vacua. 
The much larger symmetry structure
is obtained in compactifications to two dimensions, 
and give rise to 24 dimensional lattices. 
Ref. \cite{ffmt} alluded to possible similarities with the 
Massive Spectrum boson--fermion Degeneracy Symmetry [MSDS]
\cite{msds},
which arises from a basic Jacobi--like identity
in 24 dimensions. 
The compactifications to two dimensions are connected to 
24 dimensional lattices and the symmetries of those are
related to the so called moonshine symmetries. 
In two dimensions the spectral flow operator 
that induces the spinor--vector duality and the 
the twist operators that acts on the internal coordinates
can be seen to share a common structure in that both
have four periodic right--moving fermions. One may further envision that 
under decompactification back to 4 dimensions the two spinor--vector dual 
vacua appear on the boundaries of the moduli space. This is 
reminiscent of the case when spacetime supersymmetry is broken
to $N=0$ by a Scherk--Schwarz mechanism in nine dimensions
and the supersymmetric and 
non--supersymmetric vacua appear on the boundaries of the compactified
dimension. 

In this paper we embark on a program to explore the connection 
between the moonshine symmetries and the spinor--vector duality.
We foresee that the spinor--vector duality is a tip of the iceberg, 
and that elucidation of this connection may reveal a covering 
space of large space of string compactifications facilitating
a deeper understanding of their symmetries and connections. 
In this paper we make several modest steps to initiate the
enterprise. In section \ref{svd} we review a specific  
realisation of the spinor--vector duality, which is particularly suited 
for our purpose here. In this realisation the untwisted vector bosons
corresponding to the sixteen dimensional vector bundle of the 
heterotic--string in ten dimensions, generate an $SO(8)^4$ 
gauge symmetry. This is obtained by including in the 
construction four basis vectors with four periodic 
world--sheet fermions, and enhancement to larger gauge 
symmetries is obtained from twisted sectors. A similar 
basis vector with four periodic fermions
that acts simultaneously on the gauge degrees of freedom 
and the internal coordinates produces the twisted sectors. 
The spinor-vector
duality can be then seen to arise due to a special choice of 
the Generalised GSO (GGSO) phases. In section \ref{2dclassi}
we explore a similar construction in two dimensions and classify
the symmetries that arise on the resulting 24 dimensional lattices. 
In section \ref{Niemeier} we derive representations of 
some of the Niemeier lattices in 24 dimensions in the free fermionic
formulation. Section \ref{conclu} concludes our paper. 

\section{A novel basis}\label{svd}

In this section we review the spinor vector duality in the specific 
realisation of ref. \cite{neq2}.
Construction of a consistent four dimensional heterotic--string theory
in the light--cone gauge requires 20 left--moving and 44 right--moving 
two dimensional
real fermions \cite{fff} propagating on the world--sheet torus.
The models in this construction are 
specified in terms of a set of  basis vectors $v_i,i=1,\dots,n$,
$$v_i=\left\{\alpha_i(f_1),\alpha_i(f_{2}),\alpha_i(f_{3})\dots\right\}$$
describing the transformation properties of each fermion
\begin{equation}
f_A\to -e^{i\pi\alpha_i(f_A)}\ f_A, \ , A=1,\dots,44~,
\end{equation}
when transported along the non--contractible loops of the 
one loop vacuum to vacuum amplitude. 
The basis vectors span a space $\Xi$ which consists of $2^N$ sectors that give
rise to the string spectrum. Each sector is given by
\begin{equation}
\xi = \sum N_i v_i,\ \  N_i =0,1
\end{equation}
The spectrum is truncated by a GGSO projection whose action on a
string state  $|S>$ is
\begin{equation}\label{eq:gso}
e^{i\pi v_i\cdot F_S} |S> = \delta_{S}\ \cc{S}{v_i} |S>,
\end{equation}
where $F_S$ is the fermion number operator and $\delta_{S}=\pm1$ is the
spacetime spin statistics index.
Different sets of projection coefficients $\cc{S}{v_i}=\pm1$ consistent with
modular invariance give
rise to different models. A model is defined by a set
of basis vectors $v_i,i=1,\dots,n$
and a set of $2^{N(N-1)/2}$ independent projections coefficients
$\cc{v_i}{v_j}, i>j$.
The 64 world--sheet fermions in the light--cone gauge are denoted by:
$\psi^\mu, 
\chi^i,y^i, \omega^i, i=1,\dots,6$ (real left-moving fermions)
and
$\bar{y}^i,\bar{\omega}^i, i=1,\dots,6$ (real right-moving fermions);
${\bar\psi}^j, j=1,\dots,4$; 
$\bar{\eta}^k, k=0,1,2,3$; 
$\bar{\phi}^l,
l=1,\ldots,8$ (complex right-moving fermions). The division 
of the right--moving complex fermions 
into groups of four
is obtained by introducing four basis vectors $z_{\{0,1,2,3\}}$
into the basis.
Each of the $z_i$ contains four non--overlapping periodic fermions under
the sets 
$\{{\bar\psi}^{1,\dots,4},
{\bar\eta}^{0,1,2,3},
{\bar\phi}^{1,\dots,4},
{\bar\phi}^{5,\dots,8}
\}$.
We note that our notation here deviates from the 
conventional one in the free fermion literature by renaming
${\bar\psi}^5\equiv{\bar\eta}^0$.
To illustrate the structure of the spinor--vector duality we use a 
basis $V$ of seven boundary condition basis vectors given by: 
$$
V=\{v_1,v_2,\dots,v_{7}\},
$$
where
\begin{eqnarray}
v_1=\mathbf1&=&\{\psi^\mu,\
\chi^{1,\dots,6},y^{1,\dots,6}, \omega^{1,\dots,6}| \nn\\
& & ~~~\bar{y}^{1,\dots,6},\bar{\omega}^{1,\dots,6},
\bar{\eta}^{1,2,3},
\bar{\psi}^{1,\dots,5},\bar{\phi}^{1,\dots,8}\},\nn\\
v_2=S&=&\{\psi^\mu,\chi^{1,\dots,6}\},\nn\\
v_{3}=z_1&=&\{\bar{\phi}^{1,\dots,4}\},\nn\\
v_{4}=z_2&=&\{\bar{\phi}^{5,\dots,8}\},\nn\\
v_{5}=z_3&=&\{\bar{\psi}^{1,\dots,4}\},\nn\\
v_{6}=z_0&=&\{\bar{\eta}^{0,1,2,3}\},\nn\\
v_{7}=b_1&=&\{\chi^{34},\chi^{56},y^{34},y^{56}|\bar{y}^{34},
\bar{y}^{56},\bar{\eta}^0,\bar{\eta}^{1}\}.\label{basis2}
\end{eqnarray}

The partition function of such models is of the form

\begin{eqnarray}
  Z(\tau,\bar\tau)=&& \frac{1}{\tau_2(\eta \bar\eta)^2}\frac{1}{\eta^{10} {\bar\eta}^{22}}
\frac{1}{2^7}\sum_{a,b,s,s'}\sum_{h_1,g_1}\sum_{H_I,G_I} C[^{a,s,h_1,H_I}_{b,s',g_1,G_I}]\nonumber\\
&&\theta[^{a+s}_{b+s'}]\theta[^{a+s}_{b+s'}]\theta[^{a+s+h_1}_{b+s'+g_1}]\theta[^{a+s}_{b+s'}]\nonumber\\
&&\theta[^{a}_{b}]\theta[^{a+h_1}_{b+g_1}]\theta[^{a+h_1}_{b+g_1}]\theta[^{a}_{b}]\theta[^{a+h_1}_{b+g_1}]\theta[^{a+h_1}_{b+g_1}]\nonumber\\
&&\overline\theta[^{a}_{b}]\overline\theta[^{a+h_1}_{b+g_1}]\overline\theta[^{a+h_1}_{b+g_1}]\overline\theta[^{a}_{b}]\overline\theta[^{a+h_1}_{b+g_1}]\overline\theta[^{a+h_1}_{b+g_1}]
\nonumber\\
&&\overline\theta[^{a+H_0}_{b+G_0}]\overline\theta[^{a+H_0}_{b+G_0}]\overline\theta[^{a+H_0}_{b+G_0}]\overline\theta[^{a+H_0}_{b+G_0}]\overline\theta[^{a+H_3}_{b+G_3}]\overline\theta[^{a+H_3}_{b+G_3}]\overline\theta[^{a+H_3}_{b+G_3}]\overline\theta[^{a+H_3}_{b+G_3}]\nonumber\\
&&\overline\theta[^{a+H_1}_{b+G_1}]\overline\theta[^{a+H_1}_{b+G_1}]\overline\theta[^{a+H_1}_{b+G_1}]\overline\theta[^{a+H_1}_{b+G_1}]\overline\theta[^{a+H_2}_{b+G_2}]\overline\theta[^{a+H_2}_{b+G_2}]\overline\theta[^{a+H_2}_{b+G_2}]\overline\theta[^{a+H_2}_{b+G_2}]~.
\end{eqnarray}
The phases $C[^{a,s,h_1,H_I}_{b,s',g_1,G_I}]$ can be calculated in terms of the phases $C[^{v_i}_{v_j}]$ that define the model: If we define the vectors
\begin{eqnarray}
\alpha&=&a\mathbf1+s S+h_1 b_1+\sum_I H_I z_I=\sum n_a v_a~,\nonumber\\
\alpha'&=&b\mathbf1+s' S+g_1 b_1+\sum_I G_I z_I=\sum n_b' v_b~,\nonumber
\end{eqnarray}
then
\beq
C[^{a,s,h_1,H_I}_{b,s',g_1,G_I}]=C[^{\alpha}_{\alpha'}]=\big(\delta_\alpha\big)^{\sum_a n^\prime_a -1}\,  \big(\delta_{\alpha^\prime}\big)^{\sum_a n_a -1}\, 
 e^{-\pi i\, r(\alpha)\cdot \alpha^\prime}\, 
 \prod_{a,b}  C\left[^{\mathbf{B}_a}_{\mathbf{B}_b}\right]^{n_a n^\prime_b}~,
\eeq
where $\delta_\alpha=e^{i\pi\alpha(\psi^\mu)}$ and $r(\alpha)=\frac{\alpha-[\alpha]}{2}$ is the reduction vector which takes $\alpha$ to $[\alpha]$ with the latter having all its entries in the interval $(-1,1]$.

The models generated by the basis (\ref{basis2})
preserve $N=2$ space--time supersymmetry. 
Models that break $N=2$ to $N=1$ space--time supersymmetry
are easily incorporated by introducing a second basis vector $b_2$ 
\cite{fkr}.
The second function of the second $Z_2$ basis vector $b_2$ is
to break the untwisted 
observable symmetry gauge group from $SO(12)\times SO(4)$ to
$SO(10)\times U(1)^3$. Here the spinor-vector duality is therefore
seen in terms of $SO(12)$, rather than $SO(10)$, representations.
However, since in the $N=1$ vacua the spinor--vector duality 
operates separately on each of the $N=2$ planes \cite{fkr}, 
the discussion in terms of $N=2$ representations is sufficient. 

In the models generated by the basis in eq. (\ref{basis2}) 
all the geometrical degrees of freedom $\{y^i,\omega^i\vert
{\bar y}^i,{\bar\omega}^i\},~ i=1,\cdots,3$ are grouped together. 
The remaining 8 left--moving and 32 right--moving 
world--sheet fermions are divided into five
non--overlapping groups of eight real fermions.
In the ten dimensional supersymmetric heterotic--string such a division
%
always produces either $SO(32)$ or $E_8\times E_8$ gauge groups 
\cite{lewellen1986}.
Although, naively one may expect that other gauge symmetries, 
such as $SO(8)^4$, $SO(16)^2$ or $SO(8)\times SO(24)$ may be obtained, 
the modular properties of the partition function
forbid the other possible extensions.
In terms of the $SO(8)$ characters this property follows from
the equivalence of the $8_{_V}$, $8_{_S}$ and $8_{_C}$
$SO(8)$ representations, which enables twisted 
constructions of the $SO(32)$ or $E_8\times E_8$
gauge groups. 
This phenomena appear in the models generated by the 
basis in eq. (\ref{basis2}) and will be exploited in section 
\ref{2dclassi} below. 
The basis vector $b_1$ generates a $Z_2$ projection
which breaks $N=4$ to $N=2$ space--time supersymmetry, and breaks one
of the $SO(8)$ groups to $SO(4)\times SO(4)\equiv SU(2)^4$.

The sectors contributing to the gauge group are the
$0$--sector and the 10 anti--holomorphic sets:
\beqn
G=\{& 0, \nonumber\\
    & z_0,z_1,z_2,z_3, \nonumber\\
    & z_0+z_1,z_0+z_2,z_0+z_3,z_1+z_2,z_1+z_3,z_2+z_3~~\}~.
\label{gaugesectors}
\eeqn
The $0$--sector requires two oscillators acting on the vacuum
in the right--moving sector to produce a massless state; 
the $z_j$--sectors require one oscillator; 
and the $z_i+z_j$ sectors require no oscillators.
We first discuss the $N=4$ gauge group arising prior to the
inclusion of the basis vector $b_1$, which reduces $N=4$ to
$N=2$ space--time supersymmetry. The basis vector $b_1$ does
not produce additional enhancement sectors, and therefore
merely breaks the $N=4$ gauge group to a subgroup.

The $0$--sector gauge bosons produces the gauge symmetry
\begin{equation}
\left[SO(12)\right]\times SO(8)^4
\label{neq4gg}
\end{equation}
where the $SO(12)$ group factor arises from the 12 right--moving
world-sheet fermions $\{{\bar y},{\bar\omega}\}^{1,\cdots,6}$,
which correspond to the internal lattice at the free fermionic $SO(12)$
enhanced symmetry point. 
The $SO(8)_{3,0,1,2}$ group factors arise respectively from:
${\bar\psi}^{1,\cdots,4}$, ${\bar\eta}^{0,1,2,3}$,
${\bar\phi}^{1,\cdots,4}$, ${\bar\phi}^{5,\cdots,8}$.
The notation adheres to the conventional
notation in the quasi--realistic
heterotic--string models in the free fermionic formulation
\cite{rffm}.

The additional sectors in eq. (\ref{gaugesectors}) may produce
space--time vector bosons that enhance the untwisted four dimensional
gauge symmetry. The possible enhancements depend
on the GGSO projection coefficients $\cc{z_i}{z_j}$ with $i\ne j$.
Excluding the basis vector $b_1$ all 
vacua possess $N=4$ space--times supersymmetry, which fixes
the $\cc{S}{z_i}$ phases. Hence, there may be a priori $2^6$ possibilities
for the four dimensional gauge group, some of which are repeated.
Identical manifestations of the gauge groups arise from
twisted realisation of the group generators, due to the triality property of
the $SO(8)$ group representations. This is the four dimensional
manifestation of the twisted generation of gauge groups already
observed in the ten dimensional case. 
A few of the possibilities that may arise were classified in
ref. \cite{neq2}. The same construction will be exploited in 
section \ref{2dclassi} in the analysis of compactifications 
to two dimensions. 

\subsection{A simple example of the spinor--vector duality}

The basis vector $b_1$
reduces $N=4\rightarrow N=2$ space--times supersymmetry.
The $N=4$ vacuum
with $\left[SO(12)\right]\times SO(16)\times SO(16)$ gauge group
is realised with the GGSO projection coefficient taken to be:
\beqn
& &\cc{z_0}{z_1}=~~
\cc{z_0}{z_3}=~~
\cc{z_1}{z_2}=\nonumber\\
 &-&\cc{z_0}{z_2}=
-\cc{z_1}{z_3}=
-\cc{z_2}{z_3}=-1
\label{ggso12}
\eeqn
With this set of GGSO phases the additional sectors, beyond the $0$--sector,
that produce additional space--time vector bosons are $z_2$ and $z_3$, 
whereas those from all other sectors in eq. (\ref{gaugesectors}) 
are projected out. 
The additional projection induced by the basis vector $b_1$ breaks
the gauge symmetry arising from the $0$--sector to
\beq
\left[SO(8)\times SO(4)\right]_{\cal L}
\times \left[SO(8)_3\times SO(4)\times SO(4)\right]_O\times
\left[SO(8)_1\times SO(8)_2\right]_H
\eeq
The $SO(12)$ lattice gauge symmetry in eq. (\ref{neq4gg}) is reduced to
$\left[SO(8)\times SO(4)\right]_{\cal L}$.
The observable gauge symmetry arising from the $0$--sector is
$\left[SO(8)_3\times SO(4)\times SO(4)\right]_O$,
and
$\left[SO(8)_1\times SO(8)_2\right]_H$ 
is the hidden gauge symmetry. 
Both observable and hidden sector gauge symmetries are
enhanced. The hidden gauge symmetry is enhanced to
$\left[SO(16)\right]_H$ by the additional
vector bosons arising from the sector $z_2$.
At the $N=4$ level, the additional vector bosons from the sector $z_3$ enhance
the observable $\left[SO(8)_3\times SO(8)_0\right]_O$  gauge symmetry
to $\left[SO(16)\right]_O$.
At the $N=2$ level the $b_1$ projection
reduces $\left[SO(16)\right]_O\rightarrow \left[SO(12)\times SO(4)\right]_O
\equiv\left[SO(12)\times SU(2)_0\times SU(2)_1\right]_O$.
The $N=2$ spinor--vector duality is realised 
by the exchange of the vectorial
12 representation of $SO(12)$ with the spinorial 32 representation.
This duality is illustrated by considering two different models in which
these representations are interchanged due to the choices of the
GGSO projection coefficients. 
We remark further that the choice of GGSO projection coefficients
in eq. (\ref{ggso12}) prevent the enhancement of the 
$SO(12)\times SU(2)$ gauge symmetry to $E_7$, which is the $N=2$
analog of the enhancement of $SO(10)\times U(1)$ 
to $E_6$ at the $N=1$ level. 

The first choice of the extra GGSO projection coefficients that we consider
is given by:
\beq
\cc{b_1}{1,z_0}=-\cc{b_1}{S,z_1,z_2,z_3}=-1~.
\label{vectorialphasechoice}
\eeq
This choice defines a model with 2 multiplets in the 
$(1,2_L+2_R,12,1,2,1)$
and 2 in the
$(8,2_L+2_R,1,2,1,1)$ representations of
\beq
\left[SO(8)\times SO(4)\right]_{\cal L}
\times \left[ SO(12)\times SU(2)_0\times SU(2)_1\right]_{O}\times
\left[SO(16)\right]_H~.
\label{so84so12su2su2so16}
\eeq
The sectors producing the vectorial 12 representation
of $SO(12)$ are the sectors $b_1$ and $b_1 + z_3$, where the sector $b_1$
produces the $(1,2,2)$ representation and the sectors $b_1+z_3$
produces the $(8_{_S},1,1)$ under the decomposition
\beq
\left[SO(12)\right]_O
\rightarrow \left[SO(8)\times SO(4)\right]_O\equiv
\left[SO(8)\times SU(2)\times SU(2)\right]_O~.
\label{so12so8so4}
\eeq
All other states are projected out.
In this case there are eight multiplets in the vectorial
representation of the observable $SO(12)$, which also transform as
as doublets of the observable $SU(2)_1$.

We next consider the choice of GGSO phases given by
\beq
\cc{b_1}{1,z_0,z_1}=-\cc{b_1}{S,z_2,z_3}=-1
\label{spinorialphasechoice}
\eeq
This case defines a model with 2 multiplets in the $(1,2_L+2_R,32,1,1,1)$,
and 2 in the $(1,2_L+2_R,1,1,2,16)$, representations of
the gauge group in eq. (\ref{so84so12su2su2so16}). 
The sectors producing the spinorial 32 representation
of $\left[SO(12)\right]_O$ are the sectors $b_1+z_0$ and $b_1 + z_3+ z_0$,
where the sector $b_1+z_0$
produces the $(8_{_V},2,1)$ representation and the sectors $b_1+z_3+z_0$
produces the $(8_{_C},1,2)$ under the decomposition
given in eq. (\ref{so12so8so4}). 
The sectors producing the vectorial 16 multiplet
of the hidden $SO(16)$ gauge group are the sectors 
$b_1$ and $b_1 + z_2$, where the sector $b_1$
produces the $(8_{_V},1)$ multiplet and 
the sector $b_1+z_2$ produces the $(1,8_{_C})$ multiplet
under the decomposition
$\left[SO(16)\right]_H\rightarrow
\left[SO(8)_1\times SO(8)_2\right]_H$.
The hidden 16 multiplets
transform as doublets of the observable $SU(2)_1$ group.
All other states are projected out.
In this model there are eight multiplets in the spinorial 32
representation of the observable $\left[SO(12)\right]_O$.

We note that in the first model the vectorial 12 representation of the
observable $\left[SO(12)\right]_O$ is constructed as 
$12 = (8_{_S},1,1)\oplus (1,2,2)$, 
while in the second model the spinorials are constructed as
$32 = (8_{_V},2,1)\oplus(8_{_C},1,2)$ under the decomposition
$SO(12)\rightarrow SO(8)\times SU(2)\times SU(2)$.
At the core of the construction is the
triality of the $SO(8)$ representations
$8_{_S}\leftrightarrow 8_{_V}\leftrightarrow 8_{_C}$.
This property of the $SO(8)$ representations 
reproduces the standard 
decomposition of $SO(n+m)\rightarrow SO(n)\times SO(m)$ as
$V^{n+m}=(V^n,1)\oplus(1,V^m)$,
and
$S^{n+m}=(S^n,S^m)\oplus(C^n,C^m)$,
for the vectorial and spinorial representations of $SO(n+m)$,
respectively. 
The triality of the $SO(8)$ representations
enables the twisted realisations of the GUT gauge group and representations,
which is
$SO(12)$ in the $N=2$ models, and $SO(10)$ in $N=1$ models.
This $SO(8)$ triality is the main property in the analysis
of section \ref{2dclassi}. 

The transformation between the two models,
(\ref{vectorialphasechoice}) and (\ref{spinorialphasechoice}),
is induced by the discrete GGSO phase change
\ba
\cc{b_1}{z_1}=+1~~\rightarrow~~\cc{b_1}{z_1}=-1
\label{dualitymap}
\ea
In the models utilising the basis of eq. (\ref{basis2})
the map from
sectors that produce vectorial representations of the observable $SO(12)$
group, to sectors that
produce spinorial representations is obtained by adding
the basis vector $z_0$, which is
similar to the $x$--map of refs. \cite{xmap, fkr}.
The basis vector $z_0$ therefore acts as the spectral flow operator. 
It is a generator of the right--moving $N=2$ world--sheet supersymmetry 
in the models that preserve $(2,2)$ world--sheet supersymmetry.
It is the mirror image of the basis vector $S$, which is the spectral 
flow operator on the fermionic side of the heterotic--string. 
For appropriate choice of the discrete GGSO phases either 
the vectorial states or the spinorial states are kept in the spectrum. 
The discrete phase modification in eq. (\ref{dualitymap})
induces the spinor--vector duality map in the $N=2$ model.
The role of the basis vectors $z_2$ and $z_3$ in the models of
(\ref{vectorialphasechoice}) and (\ref{spinorialphasechoice})
is to generate
the twisted realisation of the gauge symmetry enhancement
of the $SO(8)$ gauge groups arising from the null sector.
We may further represent the spinor--vector duality in an 
orbifold representation \cite{aft}, and 
translate the duality map in eq. (\ref{dualitymap}) 
to distinct choices of the toroidal background fields \cite{cfkr,ffmt}. 
Generalisation of the spectral map transformation between 
heterotic--string vacua was extended to Gepner models
in \cite{afg}.

\section{\bf $D=2$ model classification}\label{2dclassi}

In this section we extend the classification of the symmetry 
groups to the case of compactifications to two dimensions. 
We develop the formalism and perform a complete classification 
in the simpler cases and partial classification in the more complex
cases, where complexity here entails increasing number of basis vectors. 
The primary property which is exploited in our classification is 
the triality of the $SO(8)$ representations. In section 
\ref{Niemeier} we will employ an alternative method 
to construct the 24 dimensional Niemeier lattices. 
In section \ref{conclu} we will comment on the overlap 
and differences between our analysis in sections 
\ref{2dclassi} and \ref{Niemeier} and that of \ref{svd}. 

We compactify the heterotic--string to two dimensions. 
The two dimensional
free fermions in the light-cone gauge (in the usual notation
\cite{fff,rffm}) are:
$\chi^i,y^i, \omega^i, i=1,\dots,8$ (real left-moving fermions)
and
$\bar{y}^i,\bar{\omega}^i, i=1,\dots,8$ (real right-moving fermions),
${\bar\psi}^A, A=1,\dots,4$, 
$\bar{\eta}^B, B=0,1,2,3$, 
$\bar{\phi}^\alpha,
\alpha=1,\ldots,8$ (complex right-moving fermions).
The left-- and right--moving real fermions are 
combined into complex fermions as 
$\rho_i=1/\sqrt{2}(y_i+i\omega_i), ~i=1,\cdots,8$,
${\bar\rho}_i=1/\sqrt{2}({\bar y}_i+i{\bar\omega}_i), ~i=1,\cdots,4$,
${\bar\rho}_i=1/\sqrt{2}({\bar y}_i+i{\bar\omega}_i), ~i=5,\cdots,8$. 

The class of models under investigation,
is generated by a maximal set $V$ of 7 basis vectors
$$
V=\{v_1,v_2,\dots,v_{7}\},
$$
\begin{eqnarray}
v_1=\mathbf1&=&
\{
\chi^{1,\dots,8},y^{1,\dots,8}, \omega^{1,\dots,8}| \nn\\
& & ~~~\bar{y}^{1,\dots,8},\bar{\omega}^{1,\dots,8},
\bar{\eta}^{0,1,2,3},
\bar{\psi}^{1,\dots,4},\bar{\phi}^{1,\dots,8}\},\nn\\
v_2=H_L&=&\{\chi^{1,\dots,8}, y^{1,\dots,8}, \omega^{1,\dots,8}
\},\nn\\
v_{3}=z_1&=&\{\bar{\phi}^{1,\dots,4}\},\nn\\
v_{4}=z_2&=&\{\bar{\phi}^{5,\dots,8}\},\nn\\
v_{5}=z_3&=&\{\bar{\psi}^{1,\dots,4}\},\nn\\
v_{6}=z_4&=&\{\bar{\eta}^{0,1,2,3}\},\nn\\
v_{7}=z_5&=&\{\bar{y}^{1,\dots,4},\bar{\omega}^{1,\dots,4}\},
\label{basis}
\end{eqnarray}
with the corresponding matrix of one--loop GGSO projection coefficients

\begin{equation}
{\bordermatrix{
        &{\bf 1}&H_L & & z_1  &z_2&z_3 & z_4&z_5 \cr
 {\bf 1}&     -1&  -1& & +1   & +1 & +1  & +1     & +1    \cr
     H_L&     -1&  -1& & \pm1  & \pm1 &\pm1  &  \pm1     &  \pm1      \cr
        &       &    & &       &     &     &         &            \cr
     z_1&     +1&\pm1 & &  +1  &  \pm1 & \pm1  &  \pm1     &  \pm1    \cr
     z_2&     +1&\pm1 & & \pm1 &   +1 & \pm1  &  \pm1     &  \pm1    \cr
     z_3&     +1&\pm1 & & \pm1 &  \pm1 &  +1  &  \pm1     & \pm1    \cr
     z_4&     +1&\pm1 & & \pm1 &  \pm1 & \pm1  &   +1     & \pm1    \cr
     z_5&     +1&\pm1 & & \pm1 & \pm1 &\pm1   &  \pm1     &   +1   \cr}}
\label{phasesmodel1}
\end{equation}

The analysis of the models is similar to the analysis in the four dimensional
case, where we define the GGSO projections in a similar way to 
eq. (\ref{eq:gso}), with the $\delta_S$ index being $+1$ in sectors 
in which the left--moving world--sheet fermions are anti--periodic and $-1$ 
sectors in which they are periodic. With this definition of the 
GGSO projection, consistent with modular invariance, we can 
proceed to analyse the symmetry configurations. 

\subsection{Configurations}\label{symmetryconfig}

We analyse the various configurations that arise with increasingly 
larger number of basis vectors. The simplest is the set 
$\{ {\bf 1}, H_L\}$. With this set there is only one possible 
configuration with $SO(48)$ symmetry. Climbing the complexity ladder
by adding the $z_1$ basis vector produces two possible configurations
$SO(8)\times SO(40)$ and $SO(48)$. The first is obtained from the 
the untwisted vector states and the vector states
from the sector $z_1$ are projected out, whereas the second is obtained 
by retaining the states from $z_1$ in the massless spectrum. The choice
of the phase $\cc{z_1}{H_L}=\pm1$ selects between the two 
configurations. 
The next set is obtained by adding the basis vector $z_2$ yielding the 
set $\{{\bf 1}, H_L, z_1, z_2\}$. The matrix of GGSO phases is 
given by:

\begin{equation}
{\bordermatrix{
        &{\bf 1}&H_L & & z_1     &z_2   \cr
 {\bf 1}&   -1&    -1& & +1   & +1  \cr
     H_L&   -1&    -1& & \pm1   & \pm1   \cr
        &       &    & &        &       \cr
     z_1&  +1&\pm1   & & +1  & \pm1   \cr
     z_2&  +1& \pm1  & & \pm1   & +1  \cr}}
\label{phasesmodelz1z2}
\end{equation}
Only the phases above the diagonal are independent, whereas those 
on and below the diagonal are fixed by the modular invariance rules. 
Thus, in the configurations corresponding to eq. (\ref{phasesmodelz1z2})
we have a total of three independent phases or eight possible configurations. 
Naturally, there are degeneracies in the space of configurations due 
to the permutation symmetries among the $z_i$. With the basis 
corresponding to eq. (\ref{phasesmodelz1z2}) we find a total 
of four independent configurations shown in table \ref{tabz1z2}.

\begin{table}
\begin{center}
\begin{tabular}{c c c | r}
$\cc{z_1}{H_L}$ & $\cc{z_2}{H_L}$ & $\cc{z_1}{z_2}$ &
Gauge group G\\\hline
 +  &  +  &  +  &  $SO(16) \times SO(32)$\\
 +  & $-$ &  +  &  $SO(8) \times SO(40) $\\
$-$ & $-$ &  +  &  $SO(48)$              \\
$-$ & $-$ & $-$ &  $E_8 \times SO(32)$   \\
\end{tabular}
\caption{The configuration of the symmetry group with four basis vectors.}
\label{tabz1z2}
\end{center}
\end{table}

We note that with each subsequent basis set the configurations
of the smaller sets are reproduced. This is a recurring feature
of string constructions \cite{fkr} and results from some generic 
$\theta$--function identities and redistribution of the vector states
among the different sectors.  
We next add the additional basis vector $z_3$ producing a five basis 
set $\{{\bf 1}, H_L, z_1,z_2,z_3\}$. The phases matrix is given by
\begin{equation}
{\bordermatrix{
        &{\bf 1}&H_L & & z_1     &z_2  & z_3 \cr
 {\bf 1}&   -1&    -1& & +1   & +1  & +1\cr
     H_L&   -1&    -1& & \pm1   & \pm1 & \pm1  \cr
        &       &    & &        &     &   \cr
     z_1&  +1&\pm1   & & +1  & \pm1   &\pm1 \cr
     z_2&  +1& \pm1  & & \pm1   & +1  &\pm1 \cr
     z_3&  +1& \pm1  & & \pm1   & \pm1 & +1 \cr
}}
\label{phasesmodelz1z2z3}
\end{equation}
The untwisted symmetry is
$SO(8)_1\times SO(8)_2\times SO(8)_3 \times SO(24)$.
In this case there are a total of six independent phases 
producing 64 distinct possibilities. Out of those we obtain
seven distinct configurations shown in table \ref{tabz1z2z3}. 
Four of the resulting configurations are reproductions of 
previous cases and three are new. 
%
a complete analysis of all configurations has been performed
in the case with five basis vectors. 

\begin{table}
\begin{center}
\begin{tabular}{c c c c c c | r}
$\cc{z_1}{H_L}$ & $\cc{z_2}{H_L}$ & $\cc{z_3}{H_L}$ &
$\cc{z_1}{z_2}$ & $\cc{z_1}{z_3}$ & $\cc{z_2}{z_3}$ &
Gauge group G\\\hline
 +  &  +  &  +  &   +   &   +   &  +  & $SO(24) \times SO(24)$\\
 +  &  +  &  +  &   +   &   +   & $-$ & $SO(8) \times SO(16)\times SO(24) $\\
 +  &  +  & $-$ &   +   &   +   &  +  & $SO(16)\times SO(32)$  \\
$-$ & $-$ &  +  &   +   &   +   &  +  & $SO(8)\times SO(40)$   \\
$-$ & $-$ &  +  &  $-$  &   +   &  +  & $E_8 \times SO(8)\times SO(24)$  \\
$-$ & $-$ & $-$ &   +   &   +   &  +  & $SO(48)$  \\
$-$ & $-$ & $-$ &  $-$  &   +   &  +  & $E_8\times SO(32)$  \\
\end{tabular}
\caption{The configuration of the symmetry group with five basis vectors.}
\label{tabz1z2z3}
\end{center}
\end{table}

The next step is to add an additional basis vector to the set. 
The set of basis vectors is then 
 $\{{\bf 1}, H_L, z_1,z_2,z_3,z_4\}$.
The untwisted symmetry is
$SO(8)_1\times SO(8)_2\times SO(8)_3 \times SO(8)_4 \times SO(16)$.
The sectors contributing to the symmetry group are the
$0$--sector and the 11 purely anti--holomorphic sets:
\beqn
G=\{& 0, \nonumber\\
    & z_1,z_2,z_3,z_4, \nonumber\\
    & z_1+z_2,z_1+z_3,z_1+z_4,z_2+z_3,z_2+z_4,z_3+z_4, \nonumber\\
    &  {\tilde z}={\bf 1}+H_L+z_1+z_2+z_3+z_4~~\}
\label{symmetrysectors}
\eeqn
where the $0$--sector requires two oscillators acting on the vacuum
in the gauge sector; the $z_j$--sectors require one oscillator; and
the $z_i+z_j$ and ${\tilde z}$ require no oscillators.
The matrix of GGSO phases is given by 

\begin{equation}
{\bordermatrix{
        &{\bf 1}&H_L & & z_1     &z_2  & z_3 & z_4\cr
 {\bf 1}&   -1&    -1& & +1   & +1  & +1& +1 \cr
     H_L&   -1&    -1& & \pm1   & \pm1 & \pm1 & \pm1  \cr
        &       &    & &        &     &   & \cr
     z_1&  +1&\pm1   & & +1  & \pm1   &\pm1& \pm1 \cr
     z_2&  +1& \pm1  & & \pm1   & +1  &\pm1 & \pm1 \cr
     z_3&  +1& \pm1  & & \pm1   & \pm1 & +1 & \pm1 \cr
     z_4&  +1& \pm1  & & \pm1   & \pm1 & \pm1 & +1 \cr
}}
\label{phasesmodelz1z2z3z4}
\end{equation}
There are 10 independent phases in eq. (\ref{phasesmodelz1z2z3z4}) 
rendering a total of 1024 different possibilities with a complete
analysis seemingly prohibitive. For a sample of the choices
we reproduce the previous seven configurations and obtain 
six new ones. The thirteen configurations are displayed 
in table \ref{tabz1z2z3z4}. 
\afterpage{%
    \clearpage
    \begin{landscape}
        \centering 
        \begin{table}
	\begin{center}
	\begin{tabular}{c c c c c c c c c c| r}
	$\cc{z_1}{H_L}$ & $\cc{z_2}{H_L}$ & $\cc{z_3}{H_L}$ & $\cc{z_4}{H_L}$ &
	$\cc{z_1}{z_2}$ & $\cc{z_1}{z_3}$ & $\cc{z_1}{z_4}$ &
	$\cc{z_2}{z_3}$ & $\cc{z_2}{z_4}$ & $\cc{z_3}{z_4}$ &
	Gauge group G\\\hline
	 +  &  +  &  +  &  +  &  +  &  +  & +  &  +  &  +  &  +  & $E_8\times SO(32)$\\
	 +  &  +  &  +  &  +  &  +  &  +  & +  &  +  &  +  & $-$ & $SO(16) \times SO(16)\times SO(16) $\\
 	 +  &  +  &  +  &  +  &  +  &  +  & +  &  +  & $-$ & $-$ & $SO(16)\times SO(8)\times SO(8) \times SO(16)$  \\
	 +  &  +  &  +  &  +  &  +  &  +  & +  & $-$ & $-$ & $-$ & $E_8\times SO(24)\times SO(8) $  \\
	 +  &  +  &  +  &  +  & $-$ &  +  & +  & $-$ &  +  & $-$ & $SO(16)\times SO(8)\times SO(8) \times SO(8)	\times SO(8)$  \\
	 +  &  +  &  +  &  +  & $-$ & $-$ &$-$ &  +  &  +  &  +  & $SO(24)\times SO(16)\times SO(8)$  \\
	 +  &  +  &  +  &  +  &  +  & $-$ &$-$ & $-$ & $-$ &  +  & $E_8\times SO(16)\times SO(16)$  \\
	$-$ &  +  &  +  &  +  &  +  &  +  & +  &  +  &  +  &  +  & $SO(24)\times SO(24)$  \\
	 +  &  +  &  +  &  +  & $-$ & $-$ &$-$ & $-$ & $-$ & $-$ & $SO(32)\times SO(16)$  \\
	$-$ & $-$ &  +  &  +  & $-$ &  +  & +  &  +  &  +  &  +  & $E_8\times E_8\times SO(16)$  \\
	$-$ & $-$ & $-$ & $-$ &  +  &  +  & +  &  +  &  +  &  +  & $SO(48)$  \\
	$-$ & $-$ & $-$ &  +  &  +  &  +  & +  &  +  &  +  &  +  & $SO(40)\times SO(8)$  \\
	$-$ & $-$ & $-$ & $-$ & $-$ &  +  & +  &  +  &  +  & $-$ & $E_8\times E_8\times E_8$  \\
	\end{tabular}
	\caption{The configuration of the symmetry group with six basis vectors.}\label{tabz1z2z3z4}
	\end{center}
	\end{table}
    \end{landscape}
    \clearpage
}

The fifth case in table \ref{tabz1z2z3z4} is a new feature of the
basis set corresponding to eq. (\ref{phasesmodelz1z2z3z4}) as compared
to the earlier cases. In all the previous cases the symmetry was 
enhanced by one or more of the additional sectors, whereas in the case of 
the fifth row in table \ref{tabz1z2z3z4} all enhancements are projected out. 
Thus, this set affords a larger set of projectors that facilitate 
projection of all enhancements. This is a recurring feature, which
is frequently used in classification of fermionic string vacua 
in four dimensions. The last row in table \ref{tabz1z2z3z4} 
correspond to a model with $E_8^3$ symmetry, which is identified 
as one of the Niemeier lattices. 

The next and final step is to add an additional basis vector
which correspond to the set given in eq. (\ref{basis})
and the GGSO coefficients matrix in eq. (\ref{phasesmodel1}). 
The untwisted symmetry is 
$SO(8)_1\times SO(8)_2\times SO(8)_3\times SO(8)_4\times SO(8)_5\times SO(8)_6$,
corresponding to the six sets of right--moving worldsheet complex fermions
$$\Big\{ 
\{
{\bar\rho}^{1,2,3,4} \}; 
\{
{\bar\rho}^{5,6,7,8}\}; 
\{
{\bar\psi}^{1,2,3,4}\};
\{
{\bar\eta}^{0,1,2,3}\};
\{
{\bar\phi}^{1,2,3,4}\};
\{
{\bar\phi}^{5,6,7,8}\}
\Big\}.$$ 
The sectors contributing to the symmetry group
are the $0$--sector and the 21 purely anti--holomorphic sets:
\beqn
G=\big\{& 0, \nonumber\\
    & z_1,~z_2,~z_3,~z_4,~z_5,~z_6, \nonumber\\
    & z_1+z_2, z_1+z_3, z_1+z_4, z_1+z_5, z_1+z_6, \nn\\
    &          z_2+z_3, z_2+z_4, z_2+z_5, z_2+z_6, \nn\\
    &                   z_3+z_4, z_3+z_5, z_3+z_6,~~~~~~~~~\nn\\
    &                   z_4+z_5, z_4+z_6, z_5+z_6~~~~~~~~~\big\}
\label{maximalsymmetrysectors}
\eeqn
where $z_6={\bf 1}+H_L+z_1+z_2+z_3+z_4+z_5=\{{\bar\phi}^{5,6,7,8}\}.$,
Similarly, to the previous cases the $0$--sector requires
two oscillators acting on the vacuum in the right--moving 
sector to produce a massless state;
the $z_i$ sectors require one oscillator; and the 
$z_i+z_j$ with $i\ne j$ require no oscillators. 
All these cases require one oscillator acting on the
vacuum in the left--moving sector. There are 15 independent 
phases in eq. (\ref{phasesmodel1}) rendering a total of 
32768 possibilities, which requires a computerised analysis, 
and is beyond our scope here.

All the sets that we introduced so far involve non--overlapping 
periodic fermions, {\it i.e.} the product between any two non--trivial 
basis vectors is $0{\rm mod}4$. We can introduce additional basis
vectors with two overlapping right--moving periodic fermions, {\it i.e.} 
the product between the new basis vectors and two of those in in eq. 
(\ref{basis}) is 2. For example, a basis vector with $z_7=\{{\bar\rho}^{1,2}, 
{\bar\eta}^{2,3}\}\equiv1$ has $z_7\cdot z_1=z_7\cdot z_4 = 2$. 
We can further envision breaking $H_L$ into three 
corresponding basis vectors $z_0$, $z_8$ and $z_9$ with
$H_L=z_0+z_8+z_9$, and similarly introduce 
basis vectors with overlapping periodic complex fermions. 
A single mod 4 left--moving basis vector, with null assignment 
for the right--moving fermions,  produces 
a Jacobi--like factor, $V_8-S_8$, in the partition 
function, which produces $N=4$ spacetime supersymmetry in the four 
dimensional models. Such non overlapping left--moving 
basis vectors produce a product of Jacobi--like identities, 
whereas basis vectors with overlapping periodic fermions 
break this identity in a familiar way from the four dimensional
models. 
The action of the basis vectors with overlapping 
periodic fermions is reminiscent of the orbifold 
action in section \ref{svd} and combining a left--moving
action with a right--moving one will entail precisely that. 
This will alter the sharp division between the left-- and 
the right--movers and reduce the symmetry structures obtained
with the 24 dimensional lattices. This looks similar to the case
of toroidal orbifolds. Detailed analysis of these cases is 
beyond our scope here and will be reported in future work. 
What may be envisioned is that the symmetry structures
of the four dimensional models are rooted in the rich symmetry 
structures of the 24 dimensional lattices in two dimensions. 
In turn the free fermionic construction may provide a set of 
simple tools that can be used to explore the properties 
of the 24 dimensional lattices. In the next section
we derive some of the Niemeier lattices by using the
free fermionic tools.

\section{The Niemeier lattices}\label{Niemeier}
Some of the models we have already presented have the property that the 
modular invariant partition function factorizes into a left and a 
right--moving part:
\beq
Z(\tau,\bar\tau)=Z(\tau)Z(\bar\tau).
\eeq
For models based on the set $\{H_L, H_R\}=\{\mathbf1, H_R\}$, which might also include some of the $z_i$ 's given in \eqref{basis}, this will happen if the phases between $H_L$ and any other vector are chosen appropriately ($\cc{H_L}{\tiny\text{anything}}=-1$).

Within the class of models with factorized partition functions, there is a subclass of models for which $Z(\tau)$ and $Z(\bar\tau)$ are modular invariant by themselves. Particular cases of this type are models for which $Z(\tau)$ is a constant. These models display a Massive Spectrum Degeneracy Symmetry (MSDS) and have been studied in \cite{msds}. Here, we would like to focus more on the right-moving partition function, which for lattice compactifications is
\beq
Z(\bar\tau)=\frac{Z_\Lambda(\bar\tau)}{\bar\eta(\bar\tau)^{24}}~.
\eeq
$\Lambda$ is the lattice on which the right-moving bosons are compactified. 

Since $Z(\bar\tau)$ is modular invariant, $\Lambda$ must be an even, self-dual, $24$-dimensional lattice (assuming a compactification to two dimensions). There are $24$ such lattices classified by Niemeier 
\cite{NIEMEIER1973142} and they are presented in table 
\ref{tb:FFFrealizations}. 
With the exception of the Leech lattice that has no vectors of length $2$, the vectors of length $2$ of the remaining $23$ lattices belong to the root lattices of simple Lie groups. However, knowing the components is not enough by itself to fully define a Niemeier lattice. One must also describe how conjugacy classes among different components are coupled with each other. This is given in terms of certain \emph{glue vectors}. For example, the Niemeier lattice $D_{12}^2$ needs glue vectors $\{(s,v),(v,s)\}$ where $v$ and $s$ stand for the vector and the spinor conjugacy class of $D_{12}$. More details about this construction and a list of glue vectors for all the Niemeier lattices can be found in \cite{conway1998sphere,ebeling2012lattices}.

Note that these lattices have been studied extensively in the past, especially in connection with moonshine. For example, the lattice $A_1^{24}$ carries a natural representation of the monster group $M_{24}$ and the Umbral Moonshine conjecture associates a finite group and a set of vector valued mock modular forms to each of these 23 Niemeier lattices (see \cite{Cheng:2013wca} and references therein).

Lattice compactifications have an equivalent fermionic description \cite{Athanasopoulos:2016aws,Athanasopoulos2016}. The main result of this section is table \ref{tb:FFFrealizations} in which we give realizations of the Niemeier lattices in terms of free fermionic basis vectors. Note that even though many glue vectors need to be included for a description in the bosonic language, the free fermionic realizations of many of these lattices are quite succinct. This is a demonstration of the power of the free fermionic formalism for certain tasks and an example where the dictionary between bosons and fermions described in \cite{Athanasopoulos:2016aws} can be used to provide new insights.

For an example, let us look at the $D_{12}^2$ Niemeier lattice again. The straightforward way one might imagine implementing this in the free fermionic language is through the basis set $\{b_1,b_2,b_3,b_4\}$ where (remembering that the normalization is twice the usual for weight vectors):
\begin{eqnarray}
b_1&=&\{1^{12},0^{12}\}~,\\
b_2&=&\{0^{12},1^{12}\}~,\\
b_3&=&\{1^{12},2,0^{11}\}~,\\
b_4&=&\{2,0^{11},1^{12}\}~.
\end{eqnarray}
$b_1$ and $b_2$ generate the two $D_{12}$, whereas $b_3$ and $b_4$ is what one would naively write down to implement the glue vectors $(s,v)$ and $(v,s)$. On the other hand, the inclusion of $b_3$ and $b_4$ appears highly unconventional from a free fermionic model building perspective because they are not independent of $b_1$ and $b_2$ when considered mod $2$. The resolution to this paradox is to use the formula \cite{Athanasopoulos:2016aws}:
\beq\label{TrivialPhaseChanges}
\cc{\alpha+\delta}{\alpha'+\delta'}=e^{\frac12 \pi i\, \delta \cdot \alpha'}\, \cc{\alpha}{\alpha'}~,
\eeq
where $\delta$ and $\delta'$ have only even entries and $\alpha,\alpha'$ are arbitrary, to reduce $b_3$ and $b_4$ to $b_1$ and $b_2$ respectively. This also changes the phase $\cc{b_1}{b_2}$ from $1$ to $-1$, hence verifying the corresponding entry in table \ref{tb:FFFrealizations}. The same idea can be applied to fill in the rest of the table. 

We tried to give realisations of the lattices using a small number of free fermionic basis vectors. However, for certain lattices we were not able to do so and we had to use the following set of $12$ basis vectors $\{g_1, g_2, \cdots, g_{12}\}$, known as the \emph{Golay generators}:
\begin{equation}
\left(
\begin{array}{cccccccccccccccccccccccc}
 1 & 0 & 0 & 0 & 0 & 0 & 0 & 0 & 0 & 0 & 0 & 0 & 0 & 1 & 1 & 1 & 1 & 1 & 1 & 1 & 1 & 1 & 1 & 1 \\
 0 & 1 & 0 & 0 & 0 & 0 & 0 & 0 & 0 & 0 & 0 & 0 & 1 & 1 & 1 & 0 & 1 & 1 & 1 & 0 & 0 & 0 & 1 & 0 \\
 0 & 0 & 1 & 0 & 0 & 0 & 0 & 0 & 0 & 0 & 0 & 0 & 1 & 1 & 0 & 1 & 1 & 1 & 0 & 0 & 0 & 1 & 0 & 1 \\
 0 & 0 & 0 & 1 & 0 & 0 & 0 & 0 & 0 & 0 & 0 & 0 & 1 & 0 & 1 & 1 & 1 & 0 & 0 & 0 & 1 & 0 & 1 & 1 \\
 0 & 0 & 0 & 0 & 1 & 0 & 0 & 0 & 0 & 0 & 0 & 0 & 1 & 1 & 1 & 1 & 0 & 0 & 0 & 1 & 0 & 1 & 1 & 0 \\
 0 & 0 & 0 & 0 & 0 & 1 & 0 & 0 & 0 & 0 & 0 & 0 & 1 & 1 & 1 & 0 & 0 & 0 & 1 & 0 & 1 & 1 & 0 & 1 \\
 0 & 0 & 0 & 0 & 0 & 0 & 1 & 0 & 0 & 0 & 0 & 0 & 1 & 1 & 0 & 0 & 0 & 1 & 0 & 1 & 1 & 0 & 1 & 1 \\
 0 & 0 & 0 & 0 & 0 & 0 & 0 & 1 & 0 & 0 & 0 & 0 & 1 & 0 & 0 & 0 & 1 & 0 & 1 & 1 & 0 & 1 & 1 & 1 \\
 0 & 0 & 0 & 0 & 0 & 0 & 0 & 0 & 1 & 0 & 0 & 0 & 1 & 0 & 0 & 1 & 0 & 1 & 1 & 0 & 1 & 1 & 1 & 0 \\
 0 & 0 & 0 & 0 & 0 & 0 & 0 & 0 & 0 & 1 & 0 & 0 & 1 & 0 & 1 & 0 & 1 & 1 & 0 & 1 & 1 & 1 & 0 & 0 \\
 0 & 0 & 0 & 0 & 0 & 0 & 0 & 0 & 0 & 0 & 1 & 0 & 1 & 1 & 0 & 1 & 1 & 0 & 1 & 1 & 1 & 0 & 0 & 0 \\
 0 & 0 & 0 & 0 & 0 & 0 & 0 & 0 & 0 & 0 & 0 & 1 & 1 & 0 & 1 & 1 & 0 & 1 & 1 & 1 & 0 & 0 & 0 & 1 \\
\end{array}
\right)
\end{equation}

Note that we only show the right-moving components of the basis vectors here, with the understanding that the left-moving components are all zero.  For a consistent free fermionic model, (at least) the vector $H_L$ should also be added to the set to ensure that the vector $\mathbf1$ is generated. The defining phases $\cc{g_i}{g_j}$ for the models that use this set are as follows:

\begin{align}\label{CGolayFirst}
A_8^3:& \quad c=&\scalebox{.8}{$\left(\arry{rrrrrrrrrrrr}{
-1&-1&-1&-1&-1&-1&-1&-1&-1&-1&-1&-1\\
*&1&1&1&-1&-1&1&1&-1&-1&1&1\\
*&*&1&1&-1&-1&1&1&-1&-1&1&1\\
*&*&*&1&-1&-1&1&1&-1&-1&1&1\\
*&*&*&*&-1&-1&1&1&-1&-1&1&1\\
*&*&*&*&*&-1&1&1&-1&-1&1&1\\
*&*&*&*&*&*&1&1&-1&-1&1&1\\
*&*&*&*&*&*&*&1&-1&-1&1&1\\
*&*&*&*&*&*&*&*&-1&-1&1&1\\
*&*&*&*&*&*&*&*&*&-1&1&1\\
*&*&*&*&*&*&*&*&*&*&1&1\\
*&*&*&*&*&*&*&*&*&*&*&1}\right)$}\\
A_7^2D_5^2:& \quad c=&\scalebox{.8}{$\left(\arry{rrrrrrrrrrrr}{
-1 & -1 & -1 & -1 & -1 & -1 & -1 & -1 & -1 & -1 & -1 & -1 \\
 * & 1 & 1 & 1 & -1 & -1 & -1 & -1 & 1 & 1 & -1 & -1 \\
 * & * & 1 & 1 & -1 & -1 & -1 & -1 & 1 & 1 & -1 & -1 \\
 * & * & * & 1 & -1 & -1 & -1 & -1 & 1 & 1 & -1 & -1 \\
 * & * & * & * & -1 & -1 & -1 & -1 & 1 & 1 & -1 & -1 \\
 * & * & * & * & * & -1 & -1 & -1 & 1 & 1 & -1 & -1 \\
 * & * & * & * & * & * & -1 & -1 & 1 & 1 & -1 & -1 \\
 * & * & * & * & * & * & * & -1 & 1 & 1 & -1 & -1 \\
 * & * & * & * & * & * & * & * & 1 & 1 & -1 & -1 \\
 * & * & * & * & * & * & * & * & * & 1 & -1 & -1 \\
 * & * & * & * & * & * & * & * & * & * & -1 & -1 \\
 * & * & * & * & * & * & * & * & * & * & * & -1}\right)$}\\
A_{15}D_9:& \quad c=&\scalebox{.8}{$\left(\arry{rrrrrrrrrrrr}{
-1 & -1 & -1 & -1 & -1 & -1 & -1 & -1 & -1 & -1 & -1 & -1 \\
 * & 1 & 1 & 1 & 1 & 1 & 1 & 1 & 1 & 1 & -1 & -1 \\
 * & * & 1 & 1 & 1 & 1 & 1 & 1 & 1 & 1 & -1 & -1 \\
 * & * & * & 1 & 1 & 1 & 1 & 1 & 1 & 1 & -1 & -1 \\
 * & * & * & * & 1 & 1 & 1 & 1 & 1 & 1 & -1 & -1 \\
 * & * & * & * & * & 1 & 1 & 1 & 1 & 1 & -1 & -1 \\
 * & * & * & * & * & * & 1 & 1 & 1 & 1 & -1 & -1 \\
 * & * & * & * & * & * & * & 1 & 1 & 1 & -1 & -1 \\
 * & * & * & * & * & * & * & * & 1 & 1 & -1 & -1 \\
 * & * & * & * & * & * & * & * & * & 1 & -1 & -1 \\
 * & * & * & * & * & * & * & * & * & * & -1 & -1 \\
 * & * & * & * & * & * & * & * & * & * & * & -1}\right)$}\\
A_4^6:& \quad c=&\scalebox{.8}{$\left(\arry{rrrrrrrrrrrr}{
 -1 & -1 & -1 & -1 & -1 & -1 & -1 & -1 & -1 & -1 & -1 & -1 \\
 * & 1 & 1 & 1 & 1 & 1 & 1 & 1 & 1 & 1 & 1 & 1 \\
 * & * & 1 & 1 & 1 & 1 & 1 & 1 & 1 & 1 & 1 & 1 \\
 * & * & * & -1 & -1 & -1 & -1 & -1 & -1 & -1 & -1 & -1 \\
 * & * & * & * & -1 & -1 & -1 & -1 & -1 & -1 & -1 & -1 \\
 * & * & * & * & * & 1 & 1 & 1 & 1 & 1 & 1 & 1 \\
 * & * & * & * & * & * & 1 & 1 & 1 & 1 & 1 & 1 \\
 * & * & * & * & * & * & * & -1 & -1 & -1 & -1 & -1 \\
 * & * & * & * & * & * & * & * & -1 & -1 & -1 & -1 \\
 * & * & * & * & * & * & * & * & * & 1 & 1 & 1 \\
 * & * & * & * & * & * & * & * & * & * & 1 & 1 \\
 * & * & * & * & * & * & * & * & * & * & * & 1}\right)$}
\label{CGolayLast}\end{align}

Note that there are some entries in table \ref{tb:FFFrealizations} for which we were unable to provide concrete realisations. We conjecture that these cases can also be given in terms of the Golay basis vectors for certain phases. However, the fact that there are $2^{78}$ {\it a priori} different phases that are allowed by modular invariance makes it computationally difficult to verify (or disprove) this claim.

For all the models presented in table \ref{tb:FFFrealizations}, an independent check of correctness  can be performed by calculating the partition function (restricted to the right-moving sector) of the proposed free fermionic realization via the formula
\beq
Z(\bar\tau)= \sum_{\text{\tiny{sectors}}\ \alpha, \beta} \cc{\alpha}{\beta}\bth{\alpha}{\beta}
\eeq
and check that it matches the partition function of the Niemeier lattices given as \cite{Dixon:1988qd}
\beq
Z(\bar\tau)=J(\bar\tau)+24(h+1)~,
\eeq
where $h$ is the corresponding Coxeter number (given for example in \cite{conway1998sphere}) and $J(\bar\tau)$ the unique modular invariant with zero constant term (\ie \ $J(\bar\tau)=j(\bar\tau)-744$). Note in particular that the partition functions of all these lattices (and those of the corresponding free fermionic models) only differ in their constant term. The massive spectra of all the models are identical.

The fact that the modular invariant phases in the partition function can be adjusted to couple conjugacy classes among different factors is quite surprising and, to our knowledge, has not been noticed before. It demonstrates an interesting interplay between gluing lattices and modularity and allows for a deeper understanding of the spinor-vector duality: For lattice compactifications, adjusting the generalized phases couples different conjugacy classes among different factors leading to different lattices. In a similar way, for orbifold compactifications adjusting certain generalized phases leads to spinor-vector dual models.

\begin{table}
\centering{
\scalebox{1}{
\( \def\arraystretch{1.4}
\arry{| c | c |}{
\hline\hline  
\textbf{Niemeier lattice} & \multirow{2}{*}{\textbf{Free fermionic basis vector realization}}\\
\textbf{based on} &
\\\hline\hline
D_{24} &  \{1^{24}\}\\
D_{16}E_8&  \{1^{16},0^8\},\{0^{16},1^8\}\\
E_8^3 &  \{\underline{1^{8},0^8,0^8}\} \\
A_{24} &  \{(\frac13)^{23}, \frac{7}3\},\ \cc{b_1}{b_1}=e^{\frac{4\pi i}{3}}\\
D_{12}^2 &  \{1^{12},0^{12}\},\{0^{12},1^{12}\},\ \cc{b_1}{b_2}=-1 \\
A_{17}E_7 &  \{(\frac13)^{18},1^6\} \\
D_{10}E_7^2 &  \{1^{24}\}, \{1^{10},\underline{(0)(1)^6,0^7}\}\\
A_{15}D_9 &  \text{Golay set} \\
D_8^3 &    \{\underline{1^{8},0^8,0^8}\} ,\ \cc{b_i}{b_j}=-1,\ i\neq j \\
A_{12}^2 &
\begin{array}{c}
\{1^8,0^{16}\}, \{0^4,1^8,0^{12}\},\{1,0^3,1,0^4,1^4,0^3,1,0^3,1,0^3\},\\
\{1,0^3,1,0^3,1,0^4,1^4,0^3,1,0^3\},\{0^{12},1^8,0^4\},\{0^{16},1^8\}, \\
 \cc{b_3}{b_4}=-1,\ \cc{b_5}{b_6}=-1
\end{array}
\\
A_{11}D_7E_6 &  ? \\
 E_6^4&  \{\underline{(0)(1)^5,(1)(0)^5,(1)(0)^5,(1)(0)^5}\} ,\ \cc{b_2}{b_4}=-1  \\
 A_9^2D_6&  ? \\
 D_6^4&  \{0^{6},1^6,0^6,1^6\}, \{1^{6},0^6,1^6,0^6\}, \{0^{12},1^{12}\},\ \cc{b_i}{b_j}=-1,\ i\neq j \\
 A_8^3&  \text{Golay set}\\
A_7^2D_5^2 &  \text{Golay set} \\
 A_6^4&  ? \\
 A_5^4D_4&  ? \\
 D_4^6&  \{1^{24}\}, \{\underline{1^{8},0^4,0^4,0^4}\} ,\ \cc{b_2}{b_5}=-1,\ \cc{b_3}{b_4}=-1 \\
 A_4^6&  \text{Golay set} \\
 A_3^8&  ? \\
 A_2^{12}&  ? \\
A_1^{24} &  ? \\
 \mbox{Leech}&  ?
\\\hline\hline  
}
\)
}
}
\caption{ \label{tb:FFFrealizations}
Free fermionic realizations of all inequivalent Neimeier lattices. Underline means permutations of all blocks separated by comma. For example, the three basis vectors needed for $E_8^3$ are $\{1^{8},0^8,0^8\},\{0^{8},1^8,0^8\},\{0^{8},0^8,1^8\}$. Only the $24$ right-moving components of the basis vectors appear explicitly; the left-moving are understood to be zero. We also state explicitly which generalized phases in the upper triangular part of the phase matrix are not $1$. For a consistent free fermionic model, (at least) the vector $H_L$ should also be added to the set to ensure that the vector $\mathbf1$ is generated.
} 
\end{table}

\section{Conclusions}\label{conclu}

The heterotic--string models in the free fermionic 
formulation are among the most realistic string models 
constructed to date \cite{rffm}. They correspond to 
$Z_2\times Z_2$ toroidal orbifold constructions 
at special points in the moduli space 
\cite{z2z21, Athanasopoulos:2016aws}. 
Their phenomenological properties raise the possibility that the 
true string vacuum shares some of their underlying properties. 
It is therefore of immense interest to explore what those underlying 
properties are. It is of course also plausible that the 
true string vacuum does not belong to this class, and for that 
purpose other classes of interesting string vacua (see {\it e.g.} 
\cite{others})  should be investigated and their underlying 
properties explored. 

A particular sub--class of free fermionic models are those
that allow for a light extra $Z^\prime$, with its distinct 
low scale signature via a di--photon excess. The construction 
of the string model utilised the spinor--vector duality, 
which is akin to mirror symmetry \cite{mirror}. A realisation of the 
spinor--vector duality relies on the triality of the 
$SO(8)$ representations. In particular, this triality 
enables a large range of possibilities for the GGSO 
phases to produce the same symmetry groups.
The $SO(8)$ triality property is also at the core of 
the well known Jacobi identity and the ensuing spacetime
supersymmetry. 

In this paper we explored the symmetry structures of heterotic--string
vacua compactified to two dimensions. Our primary motivation is to 
seek the origin of the four dimensional spinor--vector 
duality in the symmetry structure of 24 dimensional lattices
that are obtained in the two dimensional compactifications. 
This is analogous to the MSDS symmetry which is similarly
rooted in two dimensional compactifications. We discussed 
in section \ref{svd} how the spinor--vector duality is 
rooted in the triality property of $SO(8)$ representations
and we used this triality property in \ref{2dclassi} 
to classify some of the symmetries of the two 
dimensional compactifications. Self--duality under the 
spinor--vector duality played a key role in the construction
of the $Z^\prime$ model of ref \cite{frzprime}, with its
distinctive signature via a di--photon excess \cite{frdiphoton, adfm}.
Thus, a basic property underlying the string vacua is 
tied to a phenomenological model with its distinct experimental signature. 
In section \ref{Niemeier} we derived a representation of some 
of the Niemeier lattices in the free fermionic formulation.
The properties of 24 dimensional lattices and their moonshine
symmetries are of growing in the literature \cite{mooninter}. 
In this context it is not implausible that the free fermionic
methods can add to the set of tools that can be used to explore
the underlying mathematical structures. How, and whether,
these mathematical phenomena manifest themselves in 
physical observable is the arena we will explore in
future publications.

\section{Acknowledgments}

The authors declare 
that there is no conflict of interest regarding the publication of this paper.
AEF would like to thanks the Simons Center for Geometry and Physics
and the Oxford particle theory group for hospitality. 
This work is supported in part by STFC consolidated grant 
ST/L000431/1.



\bigskip
\medskip

\bibliographystyle{unsrt}

\end{document}